\documentclass[aps,pre,twocolumn,groupedaddress,showpacs,floatfix]{revtex4}
\usepackage{graphicx}
\usepackage{dcolumn}
\usepackage{bm}
\usepackage{amssymb}
\usepackage{amsmath}
\usepackage{epsfig}

\begin{document}
\title{Number of distinct sites visited by a subdiffusive random walker}
\author{Santos Bravo Yuste$^{1}$, J. Klafter$^{2}$, and
Katja Lindenberg$^{3}$}
\affiliation{$^{(1)}$ Departamento de F\'{\i}sica, Universidad de
Extremadura, E-06071 Badajoz, Spain\\
$^{(2)}$ School of Chemistry, Tel-Aviv University, Tel-Aviv, 69978
Israel\\
$^{(3)}$ Department of Chemistry and Biochemistry 0340, and Institute
for Nonlinear Science, University of California San Diego, 9500 Gilman Drive, La Jolla, CA
92093-0340, USA}

\begin{abstract}

The asymptotic mean number of distinct sites visited by a subdiffusive
continuous time random walker in two dimensions seems not to have been
explicitly calculated anywhere in the literature.
This number has been calculated for other dimensions for only one
specific asymptotic behavior of the waiting time distribution between
steps.  We present an explicit derivation for two cases in all integer
dimensions so as to formally complete a tableaux of results.  In this
tableaux we include the dominant as well as subdominant contributions in
all integer dimensions. Other quantities that can be calculated from the
mean number of distinct sites visited are also discussed.
\end{abstract}

\maketitle
The mean number of distinct sites visited by a
random walker on a lattice at time $t$ after the start of the
walk, $S(t)$, is a quantity often used to characterize such a
walk~\cite{Montroll65,Weissbook,Hughes}, and
many other quantities can be expressed in terms of
it~\cite{BluZumKlaPRB84,PREBenichou}. It is of course
well known that $S(t)$ depends on lattice dimension and geometry,
and on the nature of the walk, that is, on the form of the waiting
time distribution $\psi(t)$ between
steps~\cite{BluZumKlaPRB84,BluKlaZuOptical,BarzykinTachiyaPRL94}.  A
walk is said to be subdiffusive if the asymptotic mean square
displacement of the walker grows sublinearly with
time, e.g., $\langle r^2(t) \rangle \propto t^\gamma$
with $0<\gamma < 1$, or $\langle r^2(t) \rangle \propto \ln^\beta t$
with $\beta>0$~\cite{Havlin,Drager}.
One way to generate such subdiffusive walks is through waiting time
distributions between steps that decay sufficiently slowly to possess no
finite moments. In particular,
\begin{equation}
\label{psiDefi1}
 \psi(t)\sim \frac{\gamma \tau^\gamma}{\Gamma(1-\gamma)} t^{-1-\gamma},\quad
t\to \infty
\end{equation}
with $0< \gamma < 1$ leads to \cite{Hughes,MetKlaPhysRep}
\begin{equation}
\langle r^2(t) \rangle \sim  \frac{\Sigma^2}{\Gamma(1+\gamma) \tau^\gamma} t^\gamma
\label{msd1}
\end{equation}
while
\begin{equation}
\label{psiDefi2}
\psi(t)\sim \frac{\beta}{A t[\ln(t/\tau)]^{\beta+1}}, \quad t\to \infty
\end{equation}
with $\beta>0$ leads to \cite{Havlin,Drager}
\begin{equation}
\langle r^2(t) \rangle \sim \Sigma^2 A \ln^\beta (t/\tau)
\label{msd2}
\end{equation}
Here $A$ is a dimensionless quantity and, in both cases,
$\tau$ is a constant with units
of time and $\Sigma^2$ is the mean square displacement of a single step.
Noting the dependence of the mean square displacements on time, we will
characterize the behavior of Eqs.~\eqref{psiDefi1} and \eqref{msd1} as
\emph{slow} and that of Eqs.~\eqref{psiDefi2} and \eqref{msd2} as
\emph{ultraslow}.

The Laplace transform of $\psi(t)$ will be denoted by $\psi(u)$
(recognizable by its argument).
 For the inverse power law distribution one has \cite{Hughes}
\begin{equation}\label{}
   \psi(u)\sim 1-(\tau u)^\gamma, \quad u\to 0
\end{equation}
and, for the logarithmic distribution,
\begin{equation}\label{}
   \psi(u)\sim 1-\frac{1}{A \ln^\beta(1/\tau u)}, \quad u\to 0 .
\end{equation}
In what follows we set $\tau$ to unity.

The probability $\chi_n(t)$ that the walker has taken exactly
$n$ steps in time $t$ is a multiple convolution over
the $\psi(t)$ most easily expressed via the relation between their
Laplace transforms~\cite{BluKlaZuOptical,BlumenKWZprl84},
\begin{equation}
\mathcal{L}\chi_n(t)\equiv \chi_n(u)=\left[\psi(u)\right]^n
\left[1-\psi(u)\right]/u.
\end{equation}
The mean number of distinct sites visited is given in terms of
$\chi_n(t)$ by
\begin{equation}\label{StSn}
    S(t)=\sum_{n=0}^\infty S_n \chi_n(t),
\end{equation}
where $S_n$ is the mean number of distinct sites visited by the random
walker in $n$ steps.  It then follows that
\begin{equation}
\begin{aligned}
\mathcal{L}\left[S(t)\right]\equiv S(u)&=\left[1-\psi(u)\right] u^{-1}
\sum_{n=0}^\infty S_n \left[\psi(u)\right]^n \label{Su}.
\end{aligned}
\end{equation}

The asymptotic behavior of $S(t)$ for large $t$ can be related via the discrete
Tauberian theorem~\cite{Hughes} to the
behavior of $S(u)$ as $u \to 0$, that is, to the behavior of
$S(u)$ as $\psi(u)\to 1^-$.  The discrete Tauberian theorem says
that the expressions
\begin{equation}
\begin{aligned}
c(z) &= \sum_{n=1}^\infty c_n z^n \sim  \left(\frac{1}{1-z}\right)^\rho
F\left(\frac{1}{1-z}\right),\quad z\to 1^-
\end{aligned}
\end{equation}
and
\begin{equation}
c_n\sim \frac{n^{\rho-1}}{\Gamma(\rho)}F(n),\quad n\to\infty
\end{equation}
are equivalent if $\rho>0$, $\left\{c_n\right\}$ is a positive monotonic
sequence, and $F$ is slowly varying function at infinity in the sense
that $F(\lambda n)/F(n) \to 1$ as $n\to \infty$ for each fixed positive $\lambda$ \cite{Hughes}.  To evaluate $S(u)$ as $u\to 0$ and use this theorem to
calculate the asymptotic behavior of $S(t)$ we require the
functional forms of $S_n$.  These are well known~\cite{Hughes}:
\begin{itemize}
\item   For $d=1$ and $n\to\infty$
\begin{equation}\label{Snd1}
S_n=\left( \frac{8n}{\pi}\right)^{1/2} \left\{
1+\frac{1}{4n}+O(n^{-2})\right\},
\end{equation}
\item For $d=2$ and $n\to\infty$
\begin{equation}\label{Snd2}
S_n=  nL(n)
\end{equation}
with the slowly-varying function at infinity, $L(n)$, given by
\begin{equation}\label{Lnd2}
L(n) =\frac{a}{\ln (b n)} \left[ 1+\frac{1-\widehat\gamma}{\ln (b
n)}+O\left(\frac{1}{\ln^2(b n)}\right)\right]
\end{equation}
and where $\widehat \gamma= 0.5792\ldots$ is Euler's constant
and $a$ and $b$ depend on lattice geometry.
In particular, for a square lattice $a=\pi$ and $b=8$.
\item For $d \ge 3$ and $n\to\infty$ the leading term is of $O(n)$ for
all $d$ but the subleading terms differ,
\begin{equation}\label{}
S_n-(1-R)n\sim \begin{cases}
a \sqrt{n},& d=3\\
a \ln n,& d=4\\
a ,& d\ge 5,
\end{cases}
\end{equation}
where the probability $R$ of return to the origin depends on
$d$, and where both $R$ and  the constant $a$ depend on lattice geometry.
In particular, for a three-dimensional simple cubic lattice
$R=0.3405\ldots$~\cite{Hughes,PREBenichou}.
\end{itemize}

Therefore, to evaluate $S(u)$ it is necessary to evaluate sums of the form [see Eq.~\eqref{Su}]
\begin{equation}
\label{sums}
\mathbb{S}[f;u]\equiv\frac{1-\psi(u)}{u}\sum_{n=0}^\infty f(n) [\psi(u)]^n
\end{equation}
with $u\to 0$ and  $f(n)\sim n^\alpha$, $f(n)\sim \ln n  $, and
$f(n)\sim nL(n)$ for $n\to\infty$. Defining
\begin{equation}
\phi\left(1/u\right)\equiv \frac{1}{1-\psi(u)}
\end{equation}
and applying the discrete Tauberian theorem one finds for $u\to 0$,
\begin{align}
 \mathbb{S} [n^\alpha;u]& \sim \frac{\Gamma(1+\alpha)\phi^\alpha}{u},\\
 \mathbb{S} [\ln n;u] &\sim  \frac{\ln\phi}{u},\\
\mathbb{S}[n L(n);u] &\sim \frac{\phi L(\phi)}{u}.
\end{align}
Using these results one finds that the asymptotic dominant and
first subdominant contributions to $S(u)$ for $u\to 0$  are given by:
\begin{equation}
S(u)\sim \frac{\sqrt{2}}{u}\left( \phi^{1/2} +\frac{1}{2}\phi^{-1/2}\right),
\quad d=1
\end{equation}
\begin{equation}
S(u) \sim  \frac{\phi}{u}L(\phi),\quad d=2
\end{equation}
and
\begin{equation}\label{}
S(u)\sim  (1-R)\frac{\phi}{u} + \begin{cases}
(a\sqrt{\pi}/2u)\phi^{1/2} ,& d=3\\[1mm]
(a/u) \ln \phi,& d=4\\[1mm]
a/u ,& d\ge 5.
\end{cases}
\end{equation}

To arrive at the time-dependent survival probabilities we need to apply
the usual continuous Tauberian theorem~\cite{Hughes,Feller} with a
specific waiting time distribution. For the inverse power law
distribution~(\ref{psiDefi1}) one has $\phi(1/u)\sim u^{-\gamma}$ for $u\to 0$, so that the following asymptotic results for $t\to \infty$ follow:
\begin{equation}
S(t) \sim  \frac{\sqrt{2}\,t^{\gamma/2}}{\Gamma(1+\gamma/2)} +\frac{t^{-\gamma/2}}{\sqrt{2}\Gamma(1-\gamma/2)} ,\quad d=1
\end{equation}
and
\begin{equation}\label{}
S(t)\sim \frac{(1-R)}{\Gamma(1+\gamma)}
t^{\gamma} + \begin{cases}
 \dfrac{a\; \sqrt{\pi}\, t^{\gamma/2}}{2\Gamma(1+\gamma/2)}
,& d=3\\
a \gamma \ln t ,& d=4\\
a ,& d\ge 5.
\end{cases}
\end{equation}
We have obtained these expressions via the straightforward application of
the procedure explained in~\cite{BlumenKWZprl84} complemented with the
discrete Tauberian theorem.
The first asymptotic term for $d=1$ was obtained in \cite{YusAcePhysica04,YusteKatjaPRE05}. On the other hand, for  $d=2$ we find the asymptotic (new) result
\begin{equation}
\begin{aligned}
S(t)&\sim \frac{t^\gamma}{\Gamma(1+\gamma)}
L\left(t^{\gamma}\right)\\
&\sim \frac{t^\gamma}{\Gamma(1+\gamma)} \frac{a}{\ln (b t^\gamma)}
\left[ 1+\frac{1-\widehat\gamma}{\ln (b
t^\gamma)}+O\left(\frac{1}{\ln^{2} b t^\gamma}\right)\right].
\label{Std2}
\end{aligned}
\end{equation}
In particular, the dominant contribution, used
in~\cite{YLRAnotransBook}, is
\begin{equation}
    S(t)\propto \frac{t^\gamma}{\ln t^\gamma}.
\end{equation}
For the logarithmic waiting time distribution~(\ref{psiDefi2}) one has $\phi(1/u)\sim A\ln^\beta(1/u)$ for $u\to 0$ so that the new asymptotic results follow:
\begin{equation}
S(t) \sim  \sqrt{2A}\ln^{\beta/2} t +\frac{1}{\sqrt{2A}}\ln^{-\beta/2}
t ,\quad d=1
\end{equation}
\begin{equation}
\begin{aligned}
S(t)&\sim A\ln^{\beta} t \; L\left(A\ln^{\beta} t\right)\\
&\sim A \ln^{\beta} t \; \frac{a}{\ln\left(bA\ln^{\beta} t\right)}\left(
1+ \frac{1-\widehat\beta}{\ln\left(bA\ln^{\beta} t\right)}\right)\\
&\propto \frac{\ln^\beta t}{\ln\left(\ln^{\beta} t\right)}, \quad d=2
\end{aligned}
\end{equation}
where we have exhibited the third line to highlight the dominant term,
and
\begin{equation}\label{}
S(t)\sim (1-R)A \ln^{\beta} t
+ \begin{cases}
(a\sqrt{A\pi}/2) \ln^{\beta/2} t, & d=3\\
a \ln(A \ln^\beta  t) ,&  d=4\\
a ,& d\ge 5.
\end{cases}
\end{equation}

A number of observations about these results are interesting.
First, we turn to the compactness of our random walks as measured
by the ratio $S(t)/V(t)$, where $V(t)\sim \langle r^2(t)\rangle^{d/2}$
is the volume explored by the random walker. For the \emph{slow} random
walk this ratio behaves asymptotically as
\begin{equation}
\frac {S(t)}{V(t)} \propto
\begin{cases}
t^0, & d=1\\
(\ln t)^{-1}, & d=2\\
t^{-\gamma/2}, & d\geq 3.
\end{cases}
\end{equation}
For $d=1$ the walk is thus compact (the walker visits every
site in the region explored).  In dimensions three or greater, the walk
is non-compact.  In $d=2$ the logarithmic decay describes a walk that
is ``marginally non-compact."
For the \emph{ultraslow} walk we have
\begin{equation}
\frac {S(t)}{V(t)} \propto
\begin{cases}
t^0, & d=1 \\
[\ln (\ln^\beta t)]^{-1}, & d=2\\
(\ln t)^{-\beta/2}, &d\geq 3.
\end{cases}
\end{equation}
While in $d=1$ the walk is again compact and in
$d=2$ it is again marginally non-compact, the decay marking
non-compact behavior is slower here, and remains ``marginal" (logarithmic)
in all dimensions above $d=2$.

A second quantity of interest that can be calculated immediately from
the number of distinct sites visited is the survival probability $P(t)$  of a
stationary target surrounded by a sea of moving particles of density
$\rho$~\cite{BluZumKlaPRB84,PREBenichou,BluKlaZuOptical}:
\begin{equation}
P(t) = e^{-\rho S(t)}.
\end{equation}
The survival probability for the slow walk is a stretched exponential in
all dimensions (with a logarithmic contribution in the exponent in two
dimensions).  Particularly interesting behavior is exhibited by the
ultraslow walk,
\begin{equation}
P(t) \sim
\begin{cases}
\exp \left(-c \ln^{\beta/2} t \right), & d=1\\
\exp\left(-c \frac {\displaystyle \ln ^\beta t}{\displaystyle
\ln(\ln^\beta t)}\right), & d=2\\
\exp \left( -c \ln^\beta t \right), & d\geq 3,
\end{cases}
\end{equation}
where $c$ is a (different) constant in each case.
Note that for $d=1$ and $\beta=2$ and for $d\ge 3$ and $\beta=1$  the decay can thus be a power law, being even slower for smaller values of $\beta$. As far as we can ascertain, this is a behavior not previously observed in the target problem.

We have thus completed the panorama of ``known" but previously not
derived results (at least in the literature that we could locate)
for the asymptotic survival probability for an inverse power law waiting
time distribution~(\ref{psiDefi1}) which leads to a \emph{slow} growth
of the mean square displacement. We have presented entirely new
asymptotic survival probability results for the waiting time
distribution~(\ref{psiDefi2}), which leads to a growth of the mean
square displacement characterized as \emph{ultraslow}.  We have
concluded with a number of observations concerning the broader impact of
these results, including the diverse
range of slow decays of the survival probability in the
target problem associated with the ultraslow stepping time distribution.

\smallskip

The research of S.B.Y. has been supported by
the Ministerio de Educaci\'on y Ciencia (Spain) through grant No.\
FIS2007-60977 (partially financed by FEDER funds). K.L. is supported
in part by the National Science Foundation under grant PHY-0354937.

\end{document}